\documentclass[useAMS,usedcolumn,usenatbib]{mn2e}   
\usepackage[dvips]{graphicx}

\newcommand{\kms}{\mbox{\,km\,s$^{-1}$}}

\newcommand{\Mpc}{\mbox{\,h$^{-1}$\,Mpc}}

\newcommand{\degree}{\mbox{$^{\circ}$}}

\title[6dFGS Tiling Algorithm]
{The Tiling Algorithm for the 6dF Galaxy Survey}

\author[Campbell et al.]
{Lachlan~Campbell$^1$, Will~Saunders$^{2,3}$ and Matthew~Colless$^1$
\vspace*{6pt} \\
$^1$Research School of Astronomy and Astrophysics, The Australian National
    University, Weston Creek, ACT 2611, Australia \\
$^2$Royal Observatory, Blackford Hill, Edinburgh, EH9 3HJ, United Kingdom \\
$^3$Anglo-Australian Observatory, PO Box 296, Epping, NSW 1710, Australia}

\date{Accepted ---. Received ---; in original form ---}

\pagerange{\pageref{firstpage}--\pageref{lastpage}}
\pubyear{2002}

\begin{document}

\maketitle

\label{firstpage}

\begin{abstract}
The Six Degree Field Galaxy Survey (6dFGS) is a spectroscopic survey of
the southern sky, which aims to provide positions and velocities of
galaxies in the nearby Universe. When completed the survey will produce
approximately 170000 redshifts and 15000 peculiar velocities. The survey
is being carried out on the Anglo Australian Observatory's (AAO) UK
Schmidt telescope, using the 6dF robotic fibre positioner and
spectrograph system. We present here the adaptive tiling algorithm
developed to place 6dFGS fields on the sky, and allocate targets to
those fields. Optimal solutions to survey field placement are generally extremely difficult to find, especially in this era of large-scale galaxy surveys, as the space of available solutions is vast (2N dimensional) and false optimal solutions abound. The 6dFGS algorithm utilises the Metropolis (simulated annealing) method to overcome this problem. By design the algorithm gives uniform completeness independent of local density, so as to result in a highly complete and uniform observed sample. The adaptive tiling achieves a sampling rate of
approximately 95\%, a variation in the sampling uniformity of less than
5\%, and an efficiency in terms of used fibres per field of greater than
90\%. We have tested whether the tiling algorithm systematically biases
the large-scale structure in the survey by studying the two-point
correlation function of mock 6dF volumes. Our analysis shows that the
constraints on fibre proximity with 6dF lead to under-estimating galaxy
clustering on small scales ($<$1\Mpc) by up to $\sim$20\%, but that the
tiling introduces no significant sampling bias at larger scales. The algorithm should be generally applicable to virtually all tiling problems, and should reach whatever optimal solution is defined by the user's own merit function.
\end{abstract}


\section{Introduction}
\label{tilingintro}

\begin{figure*}
\includegraphics[width=18cm]{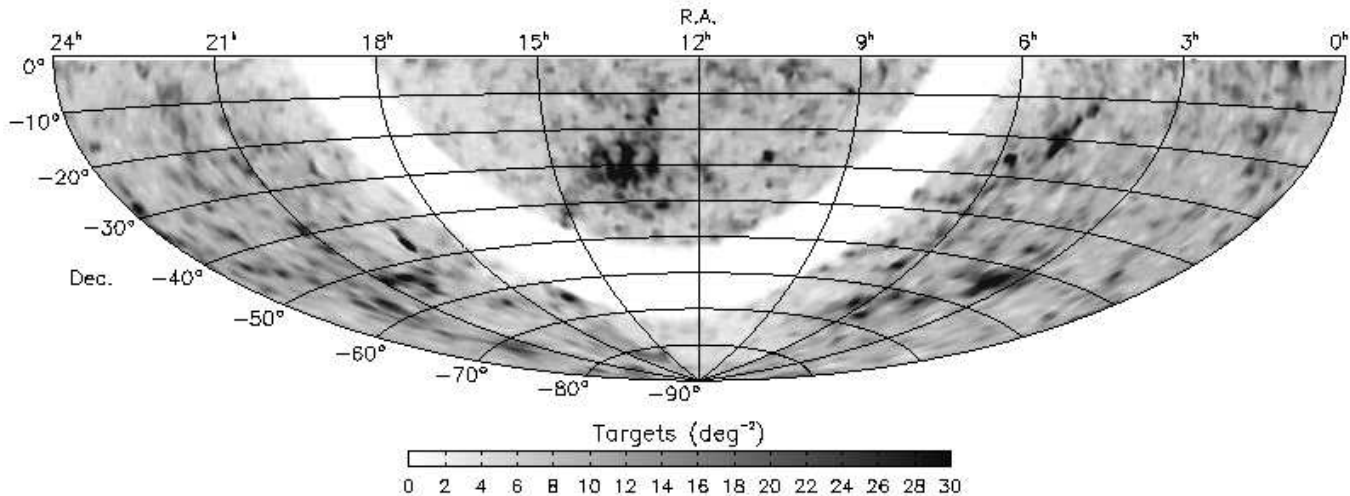}
\caption{The 6dFGS targets show strong clustering on the sky, as can be
 seen in this equal--area (Aitoff projection) greyscale map of the
 surface density of targets. As the 6dF field covers an area of
 25.5\,deg$^2$ and has up to 150 fibres, an optimal surface density
 would be approximately 6 targets per deg$^2$. The large, and spatially
 complex, density variations about this optimum illustrate one of the
 major difficulties in tiling the 6dFGS.}
\label{greyscale}
\end{figure*}

The advent of large-scale spectroscopic surveys, made possible by high
multiplex spectroscopic systems, has necessitated the development of
automated schemes for placing survey fields (`tiles') on the sky, and
allocating survey targets to those fields. Adaptive tiling schemes take
into account survey and instrument characteristics and provide efficient
and optimal tile placement and target allocation. The recently completed
2dF Galaxy Redshift Survey (2dFGRS) successfully utilized adaptive
tiling to obtain 221414 redshifts, using a 400 fibre spectrograph with a
2\degree\ field of view \citep{Colless4}. The 2dFGRS covered 2000
deg$^2$ at a median depth of $\bar{z}=0.11$. The Sloan Digital Sky
Survey (SDSS) \citep{York} aims to observe $\sim$10$^6$ targets with a
640 fibre system and a 3\degree\ field of view, and is also employing
adaptive tiling \citep{Blanton2}. The SDSS will cover
$\sim$10000\,deg$^2$ at a depth similar to the 2dFGRS.

The 6dFGS is a redshift and peculiar velocity survey that will cover the
17000 deg$^2$ of the southern sky with $|b|>10$\degree
\citep{Watson2,Saunders,Wakamatsu}. The survey is being carried out on
the AAO's Schmidt telescope, using the 6dF automated fibre positioner
and spectrograph system \citep{Parker,Watson1}. 6dF can simultaneously
observe up to 150 targets in a circular 5.7\degree\ field of view.
Survey observations are made with two different gratings for each field.
These two spectral ranges are spliced together as part of the
redshifting process, resulting in single spectra that span the range
from 3900\AA\ to 7500\AA\ , at a resolution of $R=1000$ at 5500\AA\ and a typical
signal-to-noise ratio of $S/N\sim10$.

The goals of the survey are to map the positions and velocities of
galaxies in the nearby Universe, providing new constraints on
cosmological models, and a better understanding of the local populations
of normal galaxies, radio galaxies, AGN and QSOs \citep{Saunders}. The
primary targets for the redshift survey are 113988 $K_s$-selected
galaxies from the 2MASS near-infrared sky survey (\citep{Jarrett};
{\tt{http://www.ipac.caltech.edu/2mass/releases/allsky}}) down to $K_{tot}<12.75$ and with a median redshift $\bar{z}=0.05$. The total magnitudes are estimated from the 2MASS isophotal $K_{20}$ magnitudes and surface brightness profile information \citep{Jones}. Merged with the primary sample are 16 other smaller extragalactic samples, including targets selected from the HIPASS HI
radio survey \citep{Koribalski}, the ROSAT All Sky Survey of X-ray
sources (\cite{Voges1,Voges2};
{\tt{http://heasarc.gsfc.nasa.gov/docs/rosat/ass.html}}), the IRAS Faint
Source Catalogue (\citep{Moshir};
{\tt{http://irsa.ipac.caltech.edu/IRASdocs/iras.html}}), the DENIS
near-infrared survey \citep{Epchtein}, the SuperCosmos $b_J$ and $r_F$
optical catalogues \citep{Miller}, the Hamburg-ESO QSO survey
\citep{Wisotzki} and the NVSS radio survey \citep{Condon}. In total the
survey will produce approximately 170000 redshifts.

The 6dFGS peculiar velocity survey will consist of all early-type
galaxies from the primary redshift survey sample that are sufficiently
bright to yield precise velocity dispersions. These galaxies are
observed at higher signal-to-noise ratio ($S/N>25$), in order to obtain
velocity dispersions to an accuracy of 10\%. Peculiar velocities will be
obtained using the Fundamental Plane for early-type galaxies
\citep{Djorgovski,Dressler} by combining the velocity dispersions with
the 2MASS photometry. Based on the high fraction of early-type galaxies
in the $K_s$ sample and the $S/N$ obtained in our observations to date,
we expect to measure distances and peculiar velocities for 10--15000
galaxies out to distances of at least \mbox{$cz=15000$\kms}.

Observations have so far been made for $40\%$ of the survey fields and
completion is expected mid--2005. The data is non-proprietary and an
Early Data Release for some 14000 objects can be accessed at
{\tt{http://www-wfau.roe.ac.uk/6dFGS/}}.

This paper describes the adaptive tiling algorithm developed for the
6dFGS. It is organised in the following manner: \S2 outlines the
functional requirements for the tiling algorithm and the context in
which it was developed; \S3 gives a detailed explanation of the
algorithm; \S4 outlines the process of parameter selection and
application of the algorithm to the 6dFGS catalogue; \S5 presents an
investigation of possible systematic effects introduced by the tiling,
and their impact on subsequent analyses of survey data; \S6 concludes
with a summary of the tiling algorithm and its performance.


\begin{table}
\caption{The distribution of 6dFGS targets in terms of the numbers of
 neighboring targets within the fibre-button proximity exclusion limit.
 Only 60\% of the catalogue are without close neighbours (as compared
 with $\sim$90\% in the SDSS), meaning a significant proportion have
 multiple close neighbours, the most extreme being one target with 40
 neighbours within 5.71\,arcmin.}
\begin{tabular}{crc}
\# Neighbours  & \# Targets & Sample fraction \\
 0      & 102252 & 59.2\% \\  
 1      &  43196 & 25.0\% \\ 
 2      &  15695 &  9.1\% \\
 $\ge3$ &  11604 &  6.7\% \\

\label{neighbours}
\end{tabular}
\end{table}

\section{Goals and Approach}
\label{goals}

The fundamental goals of a successful tiling algorithm are completeness,
uniformity and efficiency. Given the constraints imposed by the
instrument, the tiling algorithm should yield an arrangement of fields
that maximizes the fraction of the target sample that is observed (high
completeness) with little variation of this fraction with the position
or surface density of targets (good uniformity) and with the smallest
feasible number of fields (high efficiency).

These goals are particularly challenging for the 6dFGS. The low
redshifts of the target samples mean that even in projection on the sky
their clustering is strong, with the rms clustering per 6dF field equal to 0.64 of the mean density. Figure \ref{greyscale} shows an equal-area
(Aitoff projection) greyscale map of the surface density of targets in
the 6dFGS, illustrating the complex variations. There are also
significant instrumental constraints on fibre placement due to the large
size of the 6dF fibre buttons. These set a lower limit on the proximity
of targets that can be allocated to fibres in the same field (see \S
\ref{prox}). There is at least one neighboring target within this
proximity limit for 40.8\% of the targets in the sample (see Table
\ref{neighbours}). Despite these constraints, our requirements for the
6dF tiling algorithm were: (i)~completeness, in terms of the fraction of
total targets observed, of better than 90\%; (ii)~uniformity, in terms
of the rms variation in randomly-located 6dF fields, to better than 5\%;
(iii)~efficiency, in terms of the average fraction of fibers assigned to
targets over all fields, of at least 80\%.

The approach adopted in constructing an algorithm to achieve our goals
involves a four-stage process: (i)~the establishment of a weighting
scheme for the target galaxies to account for the relative priorities of
the target samples and to allow a balance to be set between completeness
and uniformity; (ii)~the creation of a proximity exclusion list to
account for the instrumental constraint on the closeness with which
fibres can be placed; (iii)~the initial placement of tiles and
allocation of fibres; and (iv)~the optimization of the tiling utilizing
a Metropolis algorithm \citep{Metropolis}, in order to maximize the sum
of the weights of all the allocated targets in the tiling. 

The Metropolis (simulated annealing) method was adopted because it is effective at searching very complicated parameter spaces and because it is robust against trapping
by local, rather than global, maxima \citep{Press}. While simulated annealing is expensive in terms of computation time, the entire survey is tiled at once and therefore the annealing need only be performed a few times during the life of the survey, making computation time non-critical.

Note that the tiling algorithm determines the tile locations but does
{\em not} determine the final allocation of objects to fibres in each
tile. This is because the detailed fibre configuration depends on the button and ferrule shape, fibre width, and so on; these have only secondary effect on the overall numbers of configurable targets in a field, and are in any case far too complex
and time-consuming to handle within the tiling algorithm. Final
allocations are done at the time of observation in a separate step by the
6dF {\tt configure} software, and also depend on real-time variations in the
available fibres on each of the 6dF field plates resulting from
breakages and repairs.

The tiling program was initially developed and tested using a synthetic
data set. The data came from sets of mock 6dF Galaxy Surveys,
constructed from large, high-resolution, N-body cosmological
simulations. The 6dF mock volumes have the same radial selection
function and geometrical limits as those expected for the real 6dF
survey. A full description of the method of generation, and the mock
volumes, can be found in \citealt{Cole}. The mock catalogues are
publicly available at
{\tt{http://star-www.dur.ac.uk/$\sim$cole/mocks/main.html}}. Final
testing and tuning of the algorithm was done using the 6dFGS target
catalogue.

\section{Tiling Algorithm}
\label{alg}

\subsection{Weighting Schemes}

\subsubsection{Density weighting}
\label{sd_weight}

Our original merit function was simply the overall number of targets
configured. However, this leads to a significant bias towards overdense
regions. The reason is that for any uniform level of completeness, there
are always more unallocated targets per tile in denser regions, so additional
tiles will always be placed in denser regions. In effect, the merit
function tends to equalise the number of unallocated targets per tile
area.

To get around this bias, we investigated the effect of giving each target a
weight inversely proportional to the target surface density, when smoothed
on tile scales. That is, we gave each target a density weight of
\begin{equation}
D=\left(\frac{n_6}{\langle{n_6}\rangle}\right)^{\alpha} ~~,
\end{equation}
where $n_6$ is the number of targets within the boundary of a 6dF field
centered on the target's position, $\langle{n_6}\rangle$ is the mean
number of targets per tile. With a weighting exponent $\alpha$=0 the targets are unweighted, and inverse-density weighted when $\alpha$=-1.

\subsubsection{Priority weighting}
\label{pri_weight}

Beyond the basic goals of high completeness and uniformity, we
established a target sample priority scheme to ensure the weighting
reflected the relative importance of the various samples in the survey.
The priority weight $P$ for a particular target is given by
\begin{equation}
P=\beta^p ~~,
\end{equation}
where $\beta$ is the weighting base and $p$ is the priority value
assigned to the target.

The final weight for a target is the product of its density and priority
weights, normalised to the total number of targets in the sample.

\subsection{Proximity Exclusion}
\label{prox}

The magnetic buttons of 6dF carry light-collecting prisms attached to
optical fibres that feed directly to the spectrograph slit. The buttons
are cylindrical and have a 5\,mm diameter, equivalent to 5.60\,arcmin on
the sky. This means that with a 100\,$\mu$m safety margin the minimum
separation between targets on a single tile is 5.71\,arcmin. In
optimizing the tiling, it is therefore necessary to have knowledge of
each object's proximity to other targets, in order to prevent the
allocation to the same tile of objects closer than the minimum
separation.

To achieve this, the entire catalogue of targets is searched, and a list
is created containing the number and identification of galaxies that
fall within the minimum proximity radius of a given target. This list is
consulted whenever fibres are being assigned on a given tile (see
\S\ref{ini}), and if a galaxy within the proximity exclusion zone of the
target has already been allocated to that tile, then the target is no
longer considered for assignment on that tile. The list is also used to
help prioritise the allocation of targets to tiles, as described below.

\subsection{Fibre Allocation}
\label{ini}

Two options were investigated for the initial placement of tiles: a
uniform distribution (similar to the 2dFGRS and SDSS tiling algorithms),
and placing tiles on random target positions. By doing the latter we
gain a headstart in matching the distribution of galaxies on the sky,
and tests of both methods showed that this was indeed more
computationally efficient for the 6dFGS with its high level of
clustering.

The tiling is thus initiated by choosing a target at random, placing a
tile centre at that position and assigning targets to that tile. This
process is repeated until all the pre-determined number of tiles have
been placed, with the proviso that the target chosen at random must not
already have been assigned to a tile. This approach allows a uniform
random sampling of the galaxy distribution to guide the initial
positions of the tiles.

The last step in the initial tiling is a full re-allocation of targets
to tiles. For each tile that is to have targets allocated, a list of
possible candidates---those within a tile radius of the tile centre,
2.85\degree---is created. Each candidate is given a ranking; those
targets with no neighbours within the button proximity exclusion zone
(see \S\ref{prox}) are ranked in order of increasing separation from the
tile center, since targets at the edge of the field are more likely to
be picked up by overlapping neighbouring fields. Targets with close
neighbours are ranked in order of decreasing number of neighbours, and
then increasing separation from the tile center. Candidates with close
neighbours always rank above candidates without, no matter their
separation from the tile center. The latter is to minimize
under-sampling of close pairs of targets by giving them higher priority,
in order to counteract their preferential loss due to the proximity
exclusion constraint. Once the candidate lists are complete, each tile
is assigned one target in turn, until each tile has a full complement of
targets, or has no more candidates. At all times a target can only be
allocated to a tile if it is not already allocated to another tile, and
if it is not excluded due to its proximity to a target already allocated
to the same tile.

This `democratic' allocation of targets to tiles resulted in higher
completeness and less variance in sampling than the initial method we
tested, where tiles were ordered by their number of candidates, and the
richest tile was allotted a full complement of targets before
progressing to the next richest, and so on.

\begin{figure}
\centering \includegraphics[width=8cm]{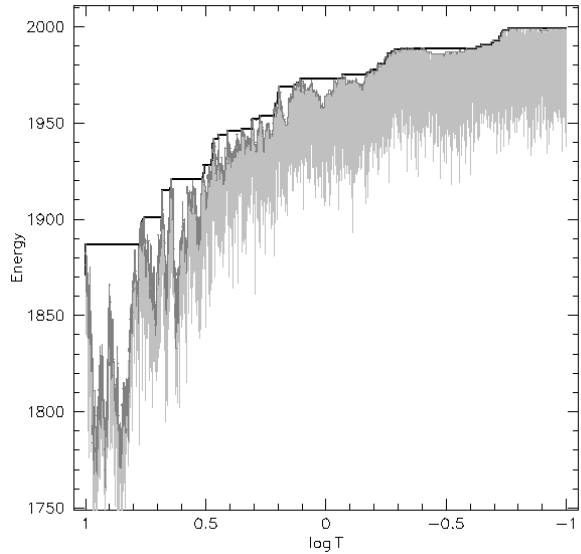}
\caption{The progress of the tiling algorithm on test data, the
horizontal axis showing the control parameter (the `temperature'), and
the vertical axis the merit function (the `energy'). The tiling
begins at an energy determined by the initial random allocation of
tiles. The tiles are then perturbed and a new configuration is accepted
(represented by the dark grey line, with the light grey line showing the
rejected configurations). As the temperature decreases the range of
changes in the accepted configurations also decreases, as the algorithm
refines its search for the optimum tiling. The best configuration at
each stage is shown by the heavy black line.}
\label{annealprog}
\end{figure}

\subsection{Optimization Process}
\label{opt}

The tiling is optimised using the Metropolis algorithm
\citep{Metropolis}, a method for simulating the natural process of
annealing. It uses a control parameter $T$ (by analogy, the
`temperature' of the tiling), and an objective function $E$ (the `energy' or merit function of the tiling), whose maximum represents the optimal tiling.
The 6dFGS tiling merit function is simply the sum of the weighted
values of all the allocated targets of a tiling. 

The annealing process is an iterative one which begins at some
predetermined temperature and at an initial value for the merit function $E_1$ computed from the initial placement of tiles and allocation of targets (see
\S\ref{ini}). We then need some
way to perturb the position of one or more tiles. This step was the
subject of extensive investigation.  Early versions perturbed the
positions of all tiles simultaneously.  However, this was found to be
grossly inefficient, because almost all such global perturbations are
unfavourable as a solution is approached. We therefore switched to
perturbing a small subset of the tiles. It was found that to randomly
select and arbitrarily reposition a single tile was also inefficient, because virtually all such individual repositionings are
unfavourable. Therefore, the tile movement was selected from a 2D Gaussian,
with rms 10\% of the tile width in each of RA and Dec. This increased run
speed to give feasible timescales, but the tiling
configuration tended to get stuck in local maxima, where no individual tile
adjustment improved the yield. A change was then made so that in 50\% of cases, all tiles within a radius of 3 tile diameters of the randomly selected tile were perturbed together, with the perturbation falling off as a Gaussian with scale length 1 tile diameter. This gave both acceptable run
times and acceptable solutions.

Following a pertubation, all nearby tiles (defined as tiles within the circle of
influence of the perturbation, with a safety margin of a degree) then have
targets reallocated. Reallocation for all tiles was neither necessary nor
computationally feasible. After
this re-allocation the merit function of the new tiling $E_2$ is computed, and
it is adopted as the current tiling with probability
\begin{equation}
 P(E_2|E_1) = \Bigg\{
\begin{array}{cr}
1                           & E_2 \geq E_1 \\
                            &              \\
exp\left[(E_2-E_1)/T\right] & E_2<E_1.
\end{array}
\end{equation}
Hence, more successful (higher energy) tilings are always accepted,
while the chances of a less successful (lower energy) tiling being
accepted decrease exponentially with the difference in the merit function, scaled by
the temperature.

After each iteration the temperature is decreased, meaning the
probability of accepting a tiling with a lower energy than the previous
one decreases as the annealing progresses. The possibly large backward
steps acceptable at the initial stages of the process are replaced by
finer changes as the tiling approaches its optimal configuration (see
Fig.\ref{annealprog}). This continues until some predetermined final
temperature, or all the targets have been allocated, whichever comes
first. The final tiling is the highest-energy tiling that occurs during
the whole course of the optimisation process.

\begin{figure*}
\includegraphics[width=18cm]{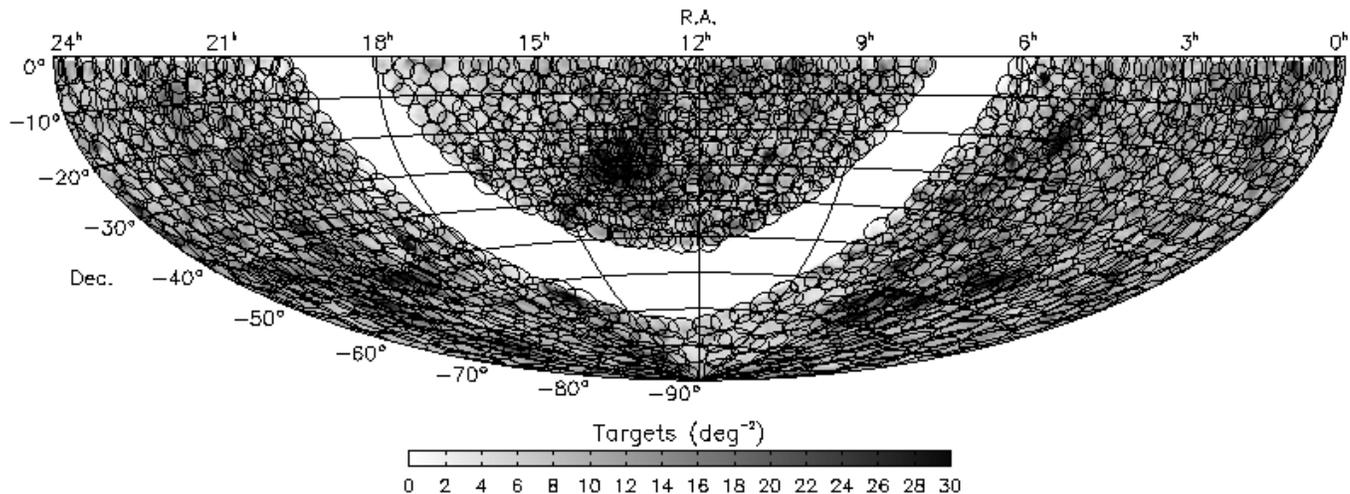}
\caption{By superimposing the fields from a tiling on the target surface density map, we can see how the tiling algorithm results in a proportional coverage in terms of
surface density, and yet still provides complete coverage of the survey
volume. Hence the adaptive tiling can achieve high completeness, as
well as highly uniform completeness.}
\label{sd_flds}
\end{figure*}

\section{Application of the Algorithm}
\label{app}

Initial survey observations were begun in a strip of the sky covering 0--360\degree R.A. and -23\degree to -42\degree in Dec., the first of the three strips of the sky selected in the survey observing strategy. These observations were made without the aid of a tiling algorithm and based on a provisional catalogue. Upon completion of the algorithm the strip was tiled, with the 50 fields already observed being included in the tiling as fixed fields. The entire survey was then tiled with the completion of the full catalogue. The algorithm parameter values used had been refined through testing upon the mock volumes and the initial Dec. strip. The tiling will continue to be an ongoing process during
the life of the survey in order to accommodate changes in strategy or circumstance. Such a circumstance arose when it became apparent in the second year of operations that inefficiencies, particularly in the early stages of the survey, required a retiling with revised tile and fibre numbers.

\begin{table}
\caption{The completeness levels for various weighting schemes using
different maximum fibre numbers per 6dF tile. The $\alpha$-value
represents the weighting exponent for the surface density weighting ,
with $\alpha=0$ corresponding to uniform weights, and $\alpha=-1$
corresponding to proportional weights (see \S\ref{sd_weight}). The
$\beta$-value is the base for the priority weighting, with $\beta=1$
meaning no priority weighting, and $\beta=2$ meaning a difference of +1
in priority makes a target twice as likely to be selected (see
\S\ref{pri_weight}). The fibre numbers were based on what we could
reasonably expect to be able to use on average, taking into account
mechanical constraints and attrition.}
\begin{tabular}{@{}cccccc}
Weighting & Priority & \multicolumn{3}{c}{Completeness} \\ 
scheme & & 125 fibres & 130 fibres & 135 fibres \\
           & 8 &94.0\% &95.1\% &95.1\% \\
$\alpha=0$ & 6 &95.8\% &97.1\% &97.1\% \\
$\beta=1$  & 5 &86.7\% &88.1\% &89.0\% \\
           & 4 &97.2\% &98.3\% &98.8\% \\
Total & & 94.5\% & 95.6\% & 95.7\% \\
&&&&\\
           & 8 &94.9\% &95.6\% &95.9\% \\
$\alpha=0$ & 6 &93.3\% &94.8\% &95.8\% \\
$\beta=2$  & 5 &83.8\% &83.0\% &85.1\% \\
           & 4 &93.7\% &96.4\% &96.7\% \\
Total & & 94.1\% & 95.1\% & 95.7\% \\
&&&&\\
            & 8 &91.2\% &92.5\% &94.3\% \\
$\alpha=-1$ & 6 &93.9\% &94.6\% &96.3\% \\
$\beta=1$   & 5 &87.8\% &88.4\% &89.3\% \\
            & 4 &97.5\% &97.6\% &98.4\% \\
Total & & 92.2\% & 93.3\% & 94.9\% \\
&&&&\\
            & 8 &92.6\% &93.8\% &94.7\% \\
$\alpha=-1$ & 6 &91.5\% &93.3\% &94.3\% \\
$\beta=2$   & 5 &85.2\% &84.4\% &84.6\% \\
            & 4 &95.3\% &95.6\% &95.7\% \\
Total & & 92.3\% & 93.6\% & 94.4\% \\
\label{fibres}
\end{tabular}
\end{table}

\subsection{Tile and Fibre Numbers}
\label{numbers}

We attempted to predict a reasonable number of fibres which could be
configured per field, given the high target clustering and mechanical
constraints such as fibre breakages and fibre crossings. Of the 150
fibres nominally available, 10 are normally assigned to blank sky
positions, leaving 140 for survey targets. Instrument commissioning and
the initial stages of the survey suggested we could expect to be able to
configure 135 of these 140 fibres per field. We compared tiling results
for a range of available fibres per tile (see Table \ref {fibres}), and
decided to limit fibre numbers within the algorithm to 135 per tile.
Based on this we needed $\sim 1330$ tiles to match target numbers. The
numbers of targets with neighbours within the fibre button proximity
exclusion zone (see Table \ref{neighbours}) also indicated we needed to
oversample the sky at least 1.5 times. Choosing 2 x oversampling, which
equated to 1360 tiles, gave us the best balance between potential sample
completeness and achievable tile numbers given the expected life-time of
the survey. The first full tiling of the catalogue was therefore tiled with 1360
tiles, each of which could be allocated a maximum of 135 targets.

By the beginning of the second year of the survey, however, it had
become apparent that this number of allocations was unrealisable, primarily
due to a higher than expected attrition rate of fibres. We therefore revised the maximum available number of fibres downwards to 125 per tile, and accordingly increased the total number of tiles to 1564 (1000 tiles for the revised tiling, and 564 tiles from the original tiling which had been observed).

\subsection{Annealing Schedule}

The annealing schedule, by which is meant the initial and final
temperatures and the steps between them, had to be chosen as a
compromise between efficacy and speed. The initial temperature
determines the size and frequency of negative changes to the tiling
configuration. Too large an initial temperature and tiles would be
relocated outside the survey region and be unable to return. Too low an
initial temperature and the annealing was unable to break out of locally
maximal configurations to achieve the global optimum. The minimum
temperature needed to be sufficiently small to allow the annealing to
perform to our expectations, without proving impractical in terms of
computation time. Finally, the temperature scale (the amount by which
the temperature is decreased after each iteration of the annealing)
needed to quench the tiling slowly enough to allow the annealing to
perform, but again could not be so slow that it would be computationally
infeasible. After testing, an initial temperature of 10 and a final
temperature of 0.1 were settled on. The temperature scale was chosen to
be a maximum of 1\% of the current temperature of the annealing, scaled
inversely to the number of tiles being configured. The larger the
tiling, the smaller the temperature scale, ensuring the annealing is
quenched more slowly in proportion to the complexity of the parameter
space.

\begin{figure*}
\includegraphics[width=18cm]{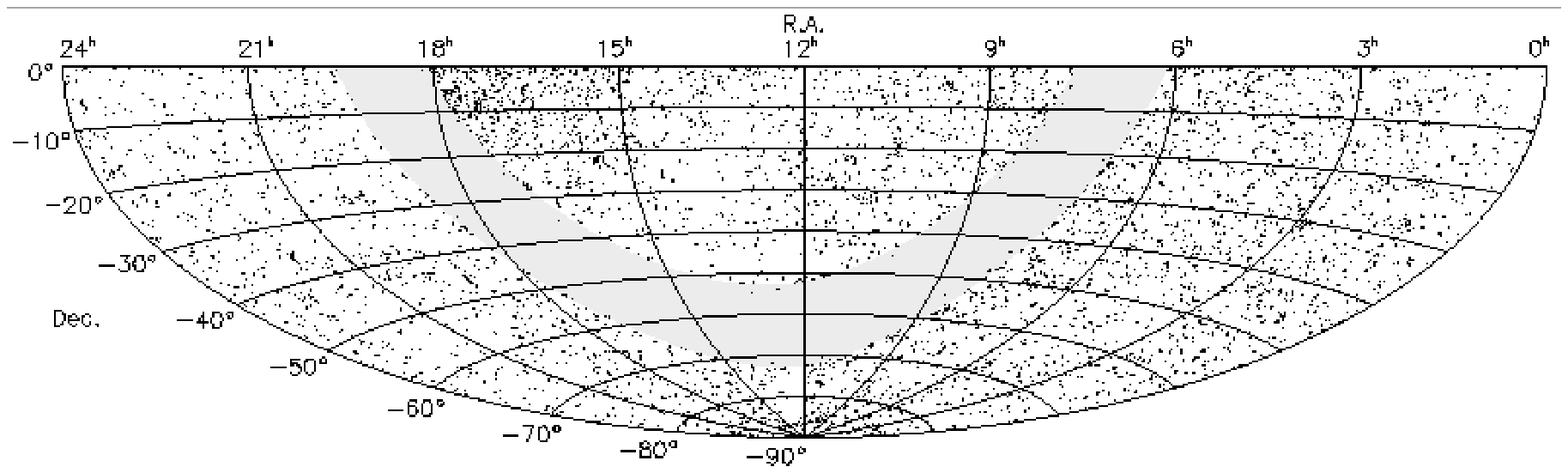}
\includegraphics[width=18cm]{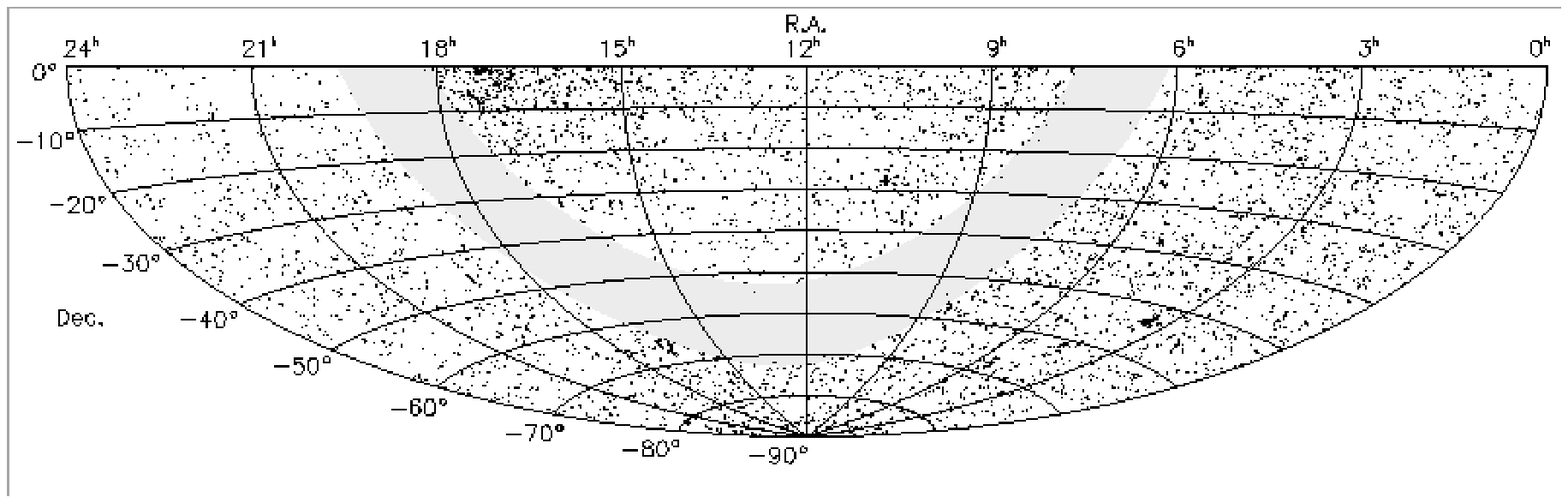}
\caption{The distribution of the survey targets not allocated to
tiles, in tilings using uniform ({\em{top}}) and proportional
({\em{bottom}}) density weights. The almost uniformly random
distribution is evidence of the success of the tiling in sampling in a
highly uniform fashion. The increased uniformity of the proportional
weighting relative to the uniform weighting is seen in the decrease in
`holes' (regions of high target density where all targets have been
allocated) in the distribution, and particularly in the better
performance along the edges of the survey.}
\label{A6ut}
\end{figure*}

\begin{figure*}
\includegraphics[width=18cm]{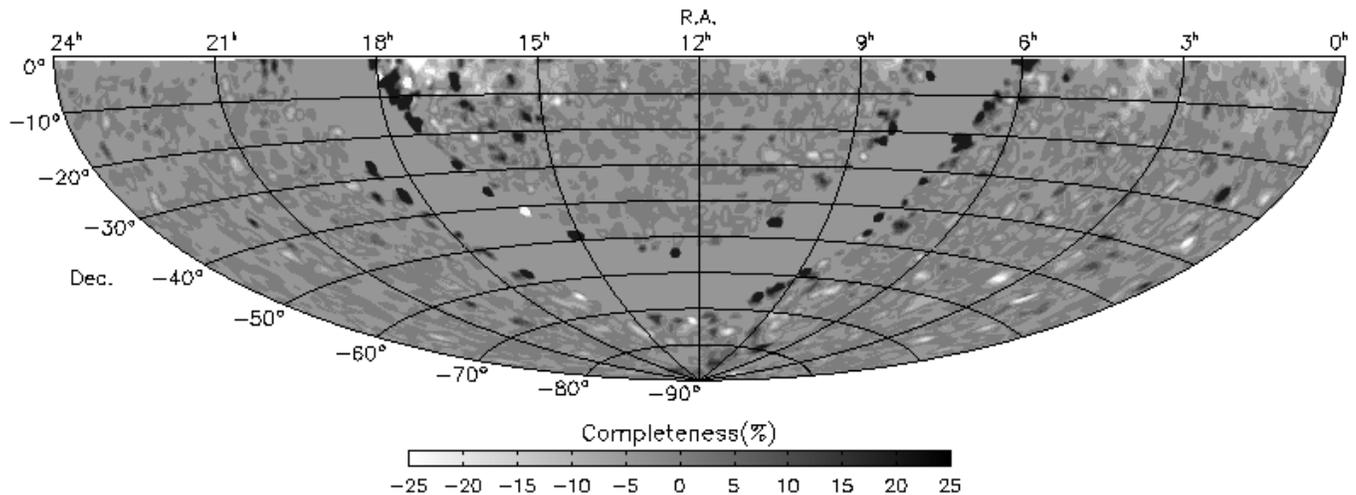}
\caption{A difference map of the completeness between the uniform and
proportional tilings of the survey: a positive difference in
favour of the proportional tiling is shown in darker colours, while a
negative difference is shown in lighter colours. The improved
performance of the proportional tiling along the edges of the survey are
obvious.}
\label{diff_cmp}
\end{figure*}

\begin{figure*}
\includegraphics[width=18cm]{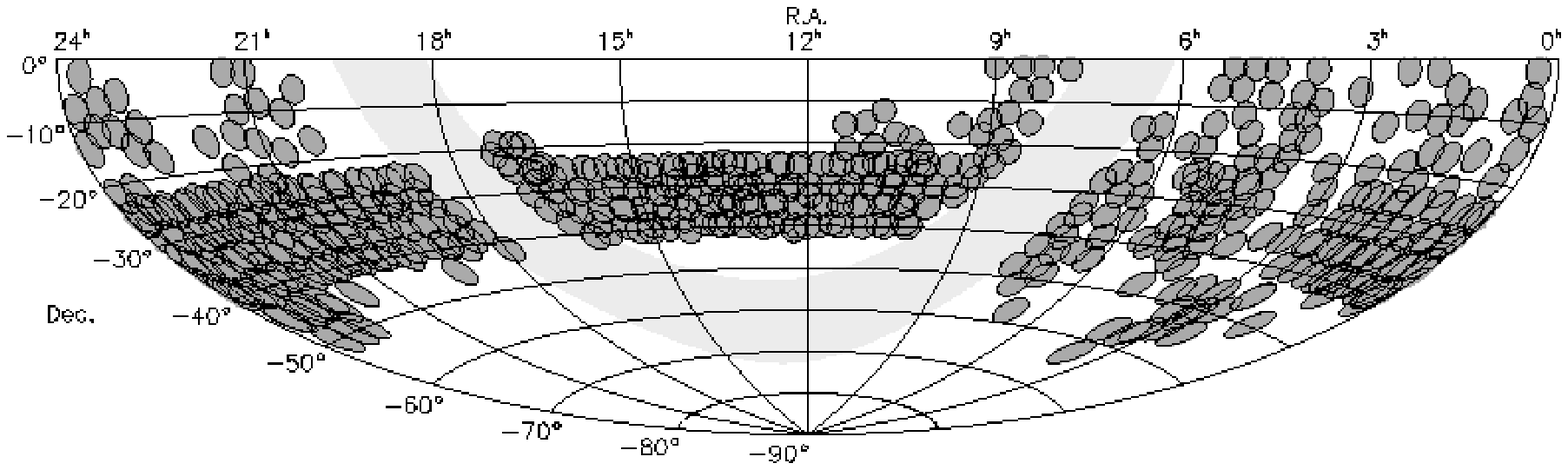}
\includegraphics[width=18cm]{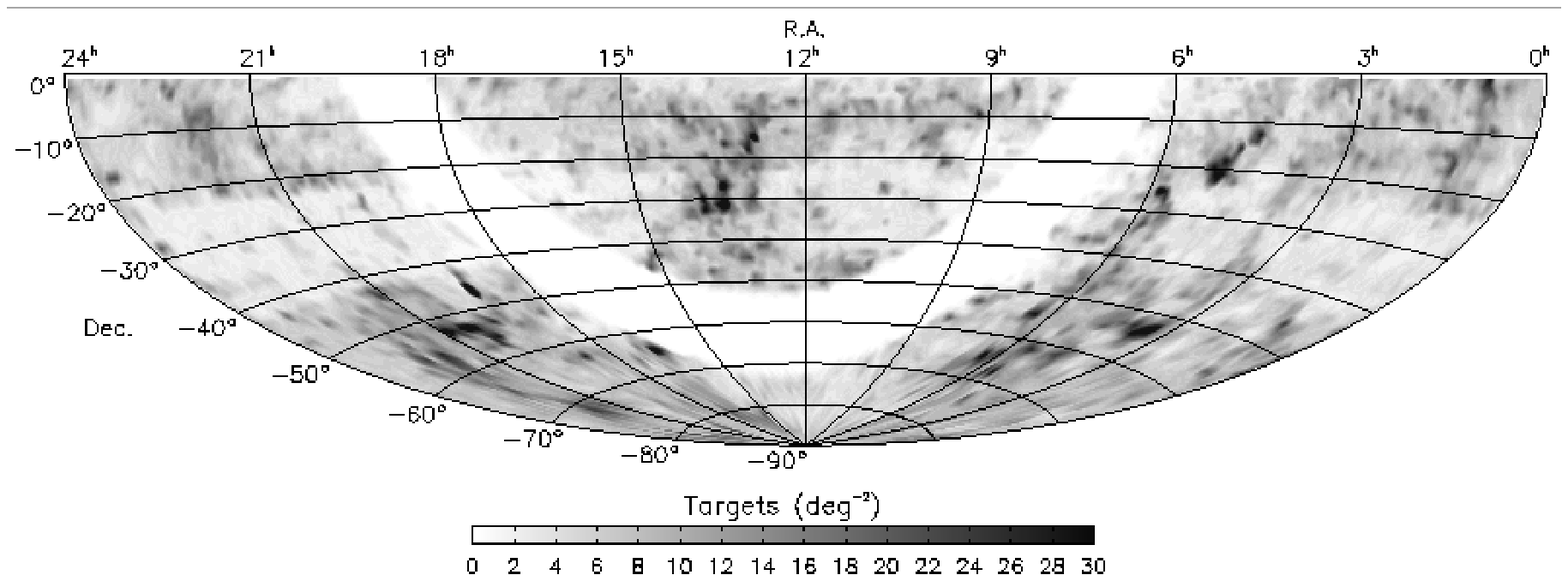}
\includegraphics[width=18cm]{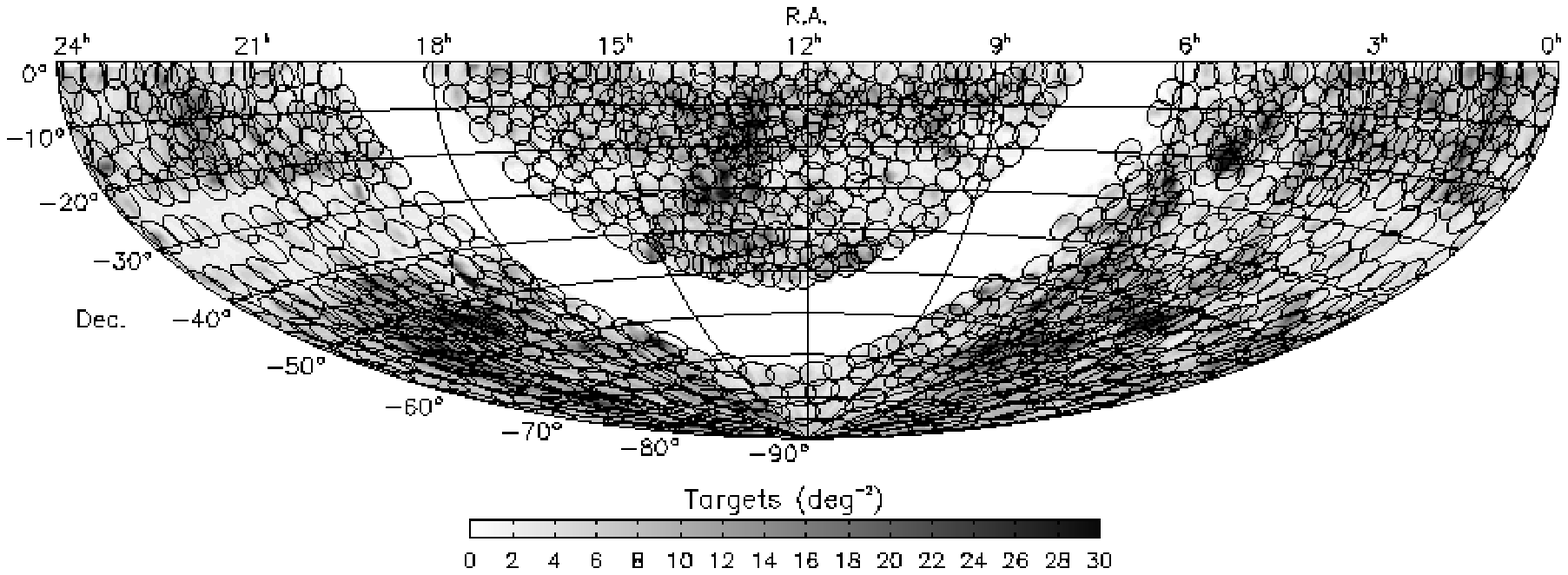}
\caption{When it became necessary to retile the survey, 564
fields from the original tiling had already been observed
({\em{top}}), most within the central declination strip of the survey
($-$23\degree\ to $-$42\degree). The targets observed successfully
within those fields were re-prioritized so as not to be included in the
new tiling, resulting in a very different distribution to be tiled
({\em{middle}}). The algorithm again provided
a tiling solution matched to the target distribution ({\em{bottom}})
which resulted in a highly complete and efficient sampling of the survey
targets.}
\label{D6tiling}
\end{figure*}

\begin{table*}
\caption{Performance statistics for the tilings of the 6dFGS 
catalogue, concentrating on the three fundamental criteria of
completeness, efficiency, and uniformity (given by the rms variation in
completeness). The survey was tiled using both a uniform
($\alpha=0$) and a proportional ($\alpha=-1$) weighting scheme. Both
tilings exceeded the performance requirements set for the algorithm.}
\begin{tabular}{@{}cccccccc}
Weighting &\multicolumn{4}{c}{Completeness}&
\multicolumn{3}{c}{Efficiency} \\
& Mean & Median & Total & RMS & Mean & Median & RMS \\ 
${\alpha=0}$  & 94.0\% & 96.0\% & 95.2\% & 3.8\% & 87.3\% & 90.4\% & 11.5\% \\
${\alpha=-1}$ & 94.5\% & 95.8\% & 94.9\% & 3.3\% & 87.0\% &  91.8\% & 13.3\% \\
\label{stats}
\end{tabular}
\end{table*}

\begin{table*}
\caption{Completeness results for individual target samples in order of
observational priority. All the tilings provide excellent completeness,
with only a small number of lower-priority samples falling below 90\%,
due to the fibre button proximity exclusion. The highest priority
targets are consistently at $\sim$95\%, indicating the success of the
priority weighting scheme.}
\begin{tabular}{@{}lrcrcc}
Sample & ID & Priority & Targets & 
\multicolumn{2}{l}{Completeness} \\
 & & & &${\alpha=0}$ &$ {\alpha=-1}$  \\ 
2MASS K$_s<12.75$           &   1 & 8 & 113988 & 95.9\% & 95.7\% \\
2MASS H$<13.05$             &   3 & 6 &   3282 & 93.7\% & 94.0\%  \\
2MASS J$<13.75$             &   4 & 6 &   2008 & 94.5\% & 94.3\%  \\
Supercosmos $r_F<15.7$      &   7 & 6 &   9199 & 95.8\% & 95.4\%  \\
Supercosmos $b_J<17.0$      &   8 & 6 &   9749 & 96.7\% & 96.5\%  \\
Shapley                     &  90 & 6 &    939 & 98.7\% & 98.2\%  \\
ROSAT All-Sky Survey        & 113 & 6 &   2913 & 95.7\% & 95.4\%  \\
HIPASS $(> 4\sigma)$        & 119 & 6 &    821 & 87.7\% & 85.8\%  \\
IRAS Faint Source Catalogue & 126 & 6 &  10707 & 96.3\% & 95.7\%  \\
Denis J$<14$                &   5 & 5 &   1505 & 91.9\% & 91.5\%  \\
Denis I$<15$                &   6 & 5 &   2017 & 74.3\% & 73.9\%  \\
2MASS AGN                   & 116 & 4 &   2132 & 95.7\% & 95.9\%  \\
Hamburg-ESO Survey          & 129 & 4 &   3539 & 96.7\% & 96.9\%  \\
NOAO-VLA Sky Survey         & 130 & 4 &   4334 & 96.3\% & 96.7\%  \\
\label{samples}
\end{tabular}
\end{table*}

\subsection{Weighting Schemes}

All of the targets in the 6dFGS catalogue have a priority based on the
relative observational importance of their particular survey sample. The
primary target sample has the highest priority of 8, while other samples
were ranked in order of their completeness requirements (lower numbers
are lower priority). Targets must have a minimum priority of 4 to be
considered in the tiling. All targets which require only serendipitous
coverage, and all successfully observed targets have priorities less
than this minimum; such targets may be included in an actual fibre
configuration (with low priority) but do not influence the tiling of the
survey (recall that the final allocation of fibres to targets is done
in a separate step at the time of observation; see \S\ref{goals}).

The priority weighting scheme uses a weighting base $\beta=2$, so that a
target with a priority one higher than another target should be twice as
likely to be allocated, based solely on its priority weight. Comparisons
of tilings with and without priority weighting typically showed an
increase in the completeness of the primary target sample (priority 8)
of up to 1\%, with lower-priority samples showing decreases of between
2\% and 5\% (Table \ref{fibres}).

When tiling the 6dFGS catalogue, the quantity $n_6$ in the density
weighting (see \S\ref{sd_weight}) is calculated from the number of
targets in the 2MASS $K_s$-selected sample alone, since this is the
primary homogeneous all-sky sample. There were two `natural' values of
the density weighting exponent $\alpha$ we could use, 0 and $-$1, which
we term uniform and proportional weighting respectively. We want
completeness, a fractional measure, to be high and uniform, but the
simplest algorithm ($\alpha=0$) just optimises on number, an absolute
measure. If we weight uniformly, then the gain for a new tile goes like
$\Delta n$ (the number of new targets acquired), which tends to maximise
overall completeness; if we weight inversely by local density
($\alpha=-1$), then we gain as $\Delta n/n_6$, which maximizes local
completeness, and so improves uniformity. In other words, uniform
density weighting optimizes global completeness, while proportional
density weighting optimizes local completeness, and hence both
completeness and uniformity. The 6dFGS catalogue can always be used to
accurately determine the true sampling as a function of position,
provided the sampling of the catalogue is not biased in terms of
spectroscopic or photometric properties of the targets. This variable
sample can then be accounted for in subsequent analyses
\citep{Colless4}. However, highly uniform sampling keeps such
corrections to a minimum; we therefore preferred, a priori, the
proportional density weighting.

\subsection{Performance analysis}

The tilings surpassed all of the goals of
completeness, uniformity, and efficiency set for the algorithm (see
Table \ref{stats}). The tiling optimization had the desired effect of
increasing tile numbers in over-dense regions, while still providing
uniform sampling of the sky and sample (see Figure \ref{sd_flds}).
The algorithm also proved to be very flexible, able to handle the highly irregular survey volume it was presented with in the revised tiling (see Figure \ref{D6tiling}).

As expected, the uniform weighting scheme resulted in the highest overall
completeness (since the tiling preferred the target-rich
densely-clustered regions), but resulted in less-uniform sampling than the proportional weighting.
As a simple form of analysis, if we display those
targets that were not allocated to fibres in the tilings, they should
appear to be uniformly randomly distributed across the sky. Figure
\ref{A6ut} shows this is the case, however the uniform tiling does show
a relatively less uniform distribution, in particular empty regions and
concentrations of targets along the edges of the survey. This edge
effect is highly apparent in Figure \ref{diff_cmp} which shows a map of
the difference in completeness between the two tilings. The dark regions
show areas where the proportional tiling resulted in higher sampling,
while the lighter shows superior performance by the uniform tiling. An
edge-avoidance effect is an understandable result of uniform tiling,
since fields placed close to the edges effectively have lower density
and hence fewer available targets. The proportional tiling's ability to
reduce this edge effect is another facet of its improved uniformity of
sampling.

Table \ref{samples} shows the completeness levels for individual target
samples. The results are excellent, with only Denis I and HIPASS sources
falling below 90\%. Denis I targets were missed due to high surface
densities, the result of stellar contamination near the Galactic Plane.
The HIPASS result can be explained by the fact that these targets are
being used to confirm the optical counterparts to radio sources, where
there are multiple possibilities in close proximity to each other.
Therefore these two samples suffer the most from the button proximity
constraint.

Close inspection of Figure \ref{A6ut} does show small concentrations of
unallocated targets, and indications of two regions of relatively poorer
sampling for both tilings. The small concentrations of unallocated
targets are primarily Denis I targets mentioned above. The North
Galactic equatorial region between $15^h$ and $18^h$ and the South
Galactic Pole however, suffer due to the combination of their low
surface densities and their proximity to the Galactic Equator. Firstly,
their low surface densities mean the initial random allocation of tiles
will sample these areas more sparsely. Secondly, tiles are unlikely to
migrate through the Equator, and hence it acts as a barrier to the free
movement of the tiles. A remedy for this would of course be to increase
tile numbers, however given the success of the tiling and the small
gains to be had, along with the constraint of a limited survey lifetime,
this was not deemed necessary.

\begin{figure}
\includegraphics[width=8.5cm]{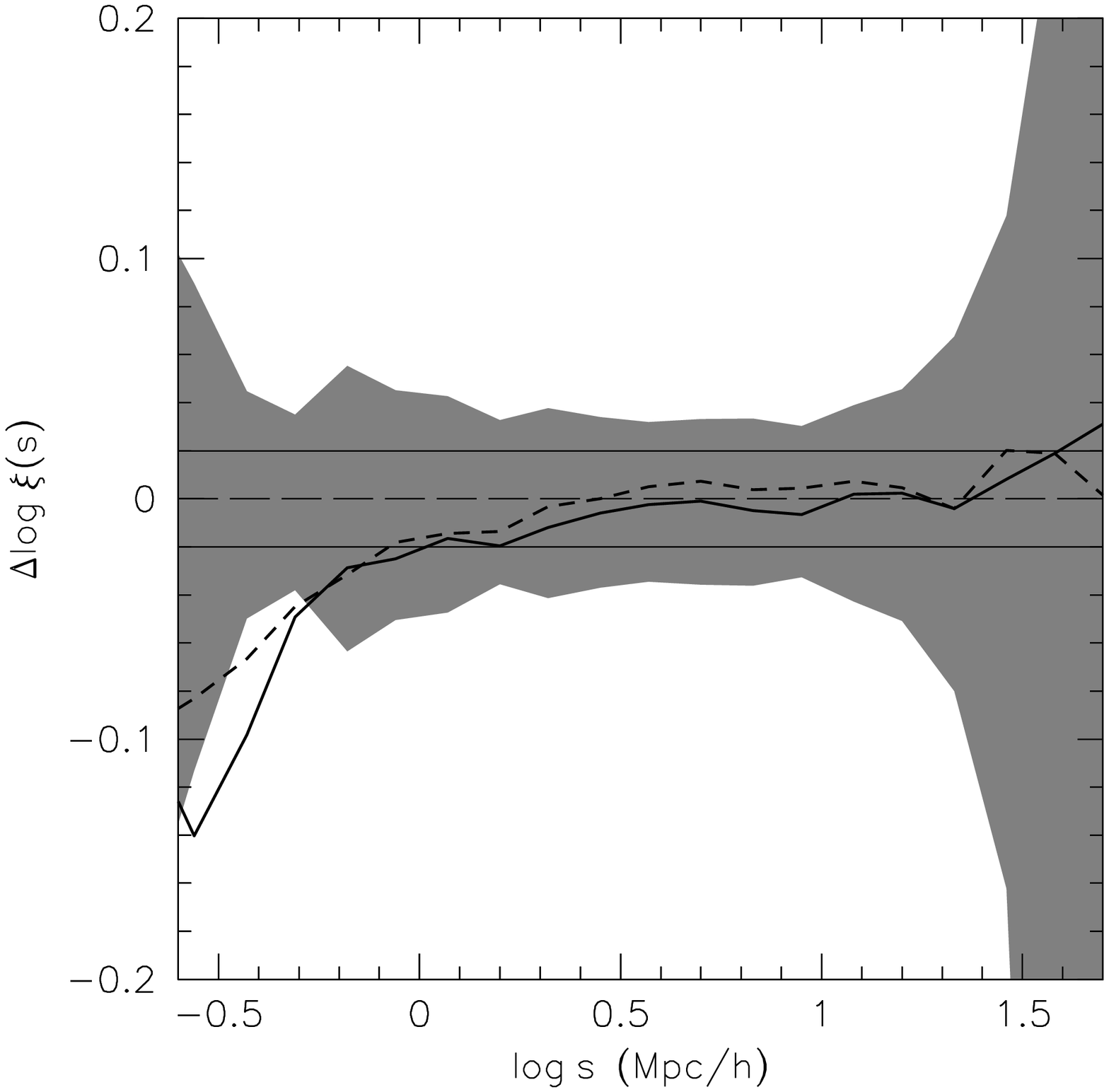}
\includegraphics[width=8.5cm]{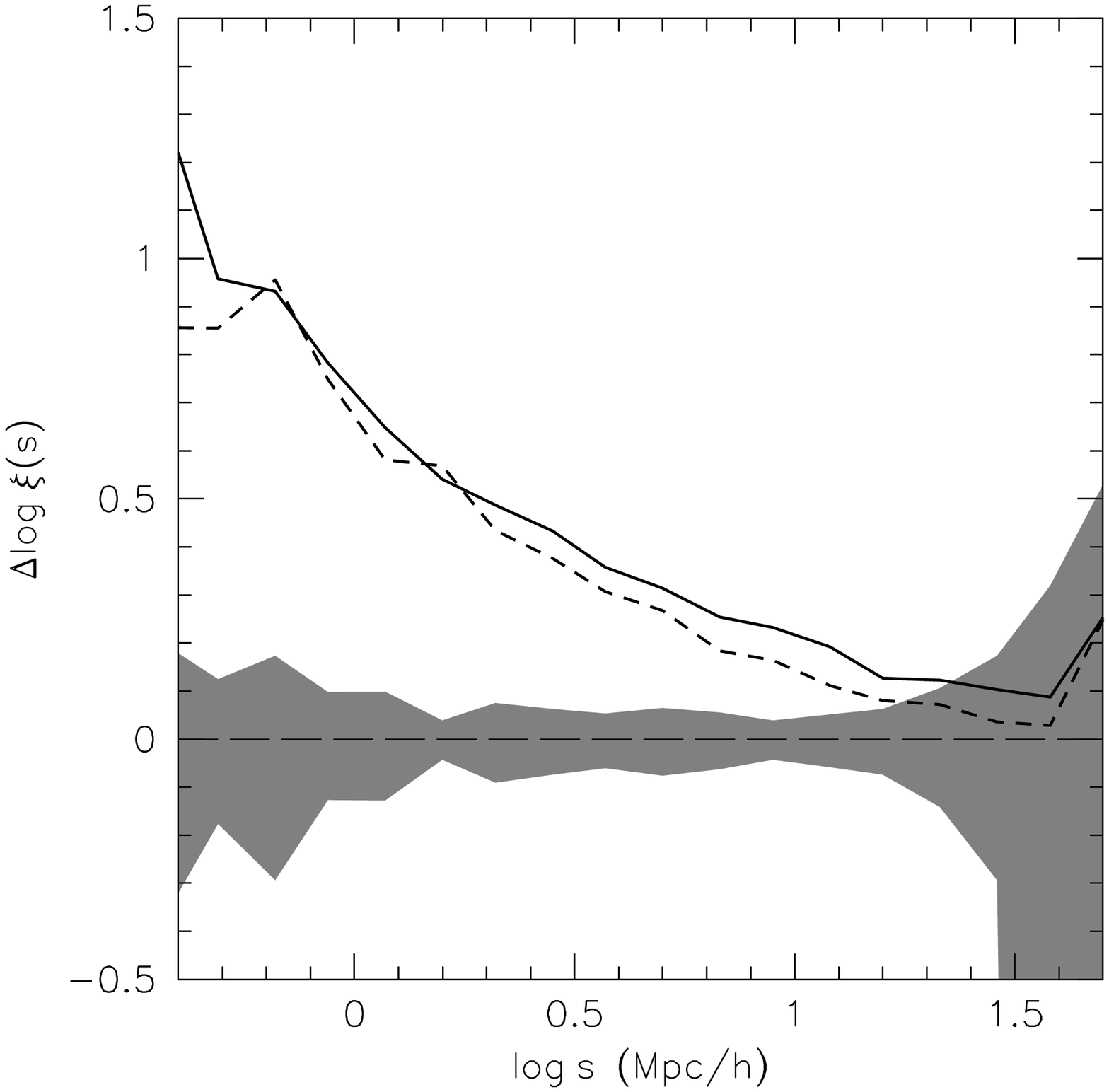}
\caption{A comparison of the mean difference in the logs of the
recovered $\xi(s)$ and the true value of $\xi(s)$, based on the
allocated ({\em{top}}) and unallocated ({\em{bottom}}) targets from
tiling 10 6dF mock volumes. The solid line represents the results from
proportional tilings, while the dashed line represents uniform tilings.
The shaded region is the $\pm 1\sigma$ variation about the mean
$\xi(s)$. The solid line either side of the zero line represents a 5\%
difference from the mean $\xi(s)$. While the recovered correlations are
consistent with the true values at large scales, there is obvious
under-estimation at smaller scales equivalent to the size of a 6dF fibre
button.}
\label{cf}
\end{figure}

\section{Systematic Effects}
\label{sys}

In order to determine the nature of any sampling biases introduced by
the tiling algorithm, and quantify their systematic effects, we compared
the two-point correlation functions of the objects in the tiled and full
samples based on mock 6dF catalogues.

We computed the correlation functions using the Landy and Szalay
estimator \citep*{Landy}. One change was made to accommodate the wide
angular coverage of the 6dFGS. The redshift space separation between two
nearby galaxies is given by
\begin{equation}
s = \sqrt{s_1^2 + s_2^2 - 2s_1s_2cos\theta}
\end{equation}
where $s_1$ and $s_2$ are the redshift space distances of the galaxies,
and $\theta$ is their angular separation on the sky. However, this
Euclidean approximation is insufficient for such a wide-angle survey as
the 6dFGS. The general formula developed by \cite*{Matsubara}, which
includes wide-angle effects and cosmological distortions, reduces, in
the case of a flat Universe, to
\begin{equation}
d = \sqrt{d_{1}^2 + d_{2}^2 - 2d_{1}d_{2}cos\theta}
\end{equation}
where $d$ is the co-moving distance of a galaxy.

The correlation function code was applied to a number of 6dF mock
volumes, and the results were consistent both with the known correlation
function of the mocks and the observed correlation functions from the
2dFGRS \citep{Hawkins} and SDSS \citep{Zehavi} surveys. Once we had
established the correlation code was working satisfactorily, we were
able to test for bias by applying it to the galaxies in 10 mock 6dF
Galaxy Surveys, and to the allocated and unallocated targets resulting
from applying the tiling algorithm to these mock surveys. Bias would
most likely appear in two forms:
\begin{enumerate}
\item The tiled sample might over- or under-represent clustered regions
of galaxies. This would distort $\xi(s)$ on the scale of a 6dF tile,
that is $\sim$6\degree, corresponding to $\sim$20\Mpc\  at the median
redshift of the survey ($z\approx0.05$).
\item The fibre proximity exclusion constraint might result in the loss
of close pairs of galaxies, distorting $\xi(s)$ on small scales. The
button size of 5\,arcmin corresponds to $\sim$0.3\Mpc\  at $z\approx0.05$.
\end{enumerate}

Figure \ref{cf} shows a comparison of the mean difference in the logs of
the recovered $\xi(s)$ and the true value of $\xi(s)$, for both
proportional (solid line) and uniform (dashed line) tilings. The shaded
region is the $\pm1\sigma$ variation about the mean $\xi(s)$ for the 10
mock surveys. The solid line either side of the zero line represents a
5\% difference from the mean $\xi(s)$. Both proportional and uniform
tilings produce estimates of $\xi(s)$ equivalent to the true value,
within the errors, at scales larger than about 1\Mpc. Even the
correlation for the unallocated targets, which exaggerates the effect of
any bias, is unaffected at scales equal to a 6dF tile and larger. This
suggests no significant sampling bias is occurring due to under or
over-representation of clustered regions of galaxies. At small scales
however the effects of the button proximity exclusion are readily
apparent. At $\sim$0.3\Mpc, the scale corresponding to a 6dF fibre
button, $\xi(s)$ is under-estimated by $\sim$20\%. This sampling bias at
small scales must therefore be taken into account in analysis of 6dFGS
data.

\section{Conclusion}
\label{conc}

Utilizing an optimization method based on simulated annealing, we have
successfully developed an adaptive tiling algorithm to optimally place
6dFGS fields on the sky, and allocate targets to those fields. The
algorithm involves a four-stage process: (i)~establishing individual
target weights based on target surface density and sample observational
priorities; (ii)~creating a database of all possible conflicts in
allocating neighbouring targets closer than the radius of a 6dF fibre
button; (iii)~creating an initial tiling by centering tiles on randomly
selected targets, and then allocating targets to those tiles in order of
decreasing numbers of neighbours and increasing separation from tile
centres; (iv)~and finally, using the Metropolis method in randomly
shifting the position of tiles, and then reallocating targets, to
maximise the objective function of the tiling and hence provide an
optimal tiling solution.

In order to maximise the uniformity of sampling of the 6dFGS targets, we
weight inversely with the surface density of 2MASS $K_s$ galaxies. Our
results showed this gave us superior uniformity when compared with a
simple uniform density weighting scheme, most noticeably in reducing the
number of targets not allocated to tiles along the edges of the survey
volume.

Despite the challenges of highly clustered targets and large fibre
buttons, tiling solutions generated using the algorithm are highly
complete and uniform, and employ an efficient use of tiles. The
tilings consistently give sampling rates of around 95\%, with
variations in the uniformity of sampling of less than 5\%. Tiles
typically have more than 90\% of their available fibres allocated to
targets. The algorithm has also proved itself highly flexible, able to perform on highly irregularly shaped distributions of targets. 

An analysis of the two-point correlation function, calculated from 6dF
mock volumes tiled with the algorithm, revealed that the constraint on
fibre proximity due to the large size of the fibre buttons produces a
significant under-sampling of close pairs of galaxies on scales of 1 \Mpc\
and smaller; on larger scales, however, the tiling algorithm does not
lead to any detectable sampling bias.

\section*{Acknowledgments}
We thank Shaun Cole for creating the 6dF mock volumes, Tom Jarrett for
all his help with the 2MASS target samples, and Idit Zehavi and Peder
Norberg for information on the SDSS and 2dFGRS correlation functions.
 
\bibliographystyle{mn2e}
\bibliography{Lbib}

\label{lastpage}

\end{document}